\documentclass[11pt,dvips]{article}

\usepackage{epsfig,times}
\usepackage{picinpar}
\usepackage{wrapfig}
\usepackage{floatflt}
\makeatletter
\renewcommand{\section}{\@startsection{section}{1}{0in}
	{0.4\baselineskip}{0.1\baselineskip}{\Large\bf}}
\renewcommand{\subsection}{\@startsection{subsection}{2}{0in}
	{0.25\baselineskip}{-\baselineskip}{\large\bf}}
\renewcommand{\subsubsection}{\@startsection{subsubsection}{3}{0in}
	{0.1\baselineskip}{-\baselineskip}{\normalsize\bf}}
\makeatother
\newcommand{\apj}{ApJ}
\newcommand{\apjl}{ApJ}
\newcommand{\apjs}{ApJS}
\newcommand{\nat}{Nature}
\newcommand{\mnras}{MNRAS}
\newcommand{\aj}{AJ}
\begin{document}

%
\thispagestyle{myheadings}
%
\markright{OG 2.1.30}
\begin{center}
%
{\LARGE \bf Measurements of Flux Limits for TeV Emission from AGNs}
\end{center}

\begin{center}
%
%
{\bf P.M.~Chadwick, K.~Lyons, T.J.L.~McComb, J.A.~Murray, K.J.~Orford,
J.L.~Osborne, S.M.~Rayner, S.E.~Shaw, and K.E.~Turver}\\
{\it Department of Physics, Rochester Building, Science Laboratories,
University of Durham, Durham, DH1~3LE, U.K.}
\end{center}

\begin{center}
{\large \bf Abstract\\}
\end{center}
\vspace{-0.5ex}
%
%
The University of Durham Mark 6 Telescope has been used to make
observations of a number of AGNs visible from the Southern Hemisphere.
Results are presented for limits to VHE gamma ray emission from 1ES
0323+022, PKS 0829+046, 1ES 1101--232, Cen A, PKS 1514--24, RXJ 10578--275
and 1ES 2316--423, both for steady emission and for outbursts on
timescales of $\sim 1$ day.

%

\vspace{1ex}

%
%

\section{Introduction}

The first AGN to be identified as a VHE source was Mrk 421 (Punch et al,
1992). Since then, Mrk 501 (Quinn et al. 1996), 1ES 2344+514 (Catanese
et al. 1998) and PKS 2155--304 (Chadwick et al. 1999a) have been
identified as VHE sources. Some of these objects have shown periods of
intense flaring activity on timescales as short as 15 minutes (Gaidos et
al. 1996, Aharonian et al. 1999).

Previous surveys of AGNs (Kerrick et al. 1995, Roberts et al. 1999,
Rowell et al. 1999) have indicated that only close X-ray selected BL
Lacs are observable at TeV energies. This is in accord both with
theoretical models of gamma ray emission from AGNs and absorption of VHE
gamma rays on the cosmic infra-red background. Recent progress in the
understanding of blazars has shown that the distinction between
radio-selected (RBL) and X-ray selected (XBL) BL Lacs is not exact and
that there is a continuum between these two extremes (Ghisellini et al.
1998). 

The Durham AGN dataset consists of observations of 10 AGNs made with the
Mark 6 telescope between 1996 and 1998. The discovery of VHE gamma rays from
PKS 2155--304 has already been reported; this is the most distant BL Lac
yet detected at these energies (Chadwick et al. 1999a). Here we describe
observations of 1ES 0323+022, PKS 0829+046, RXJ 10578--275, 1ES
1101--232, Cen A, PKS 1514--24, and 1ES 2316--423, covering a range of
classes of AGN. The typical energy threshold for these observations is $
\sim 300~{\rm to}~400 $ GeV.

\section{Observations}

The seven AGNs which are discussed in this paper comprise three XBLs,
two RBLs, one intermediate class object and one close radio galaxy (Cen
A) which has been reported previously as a VHE $\gamma$-ray source.
While RBLs are thought to be less promising as VHE $\gamma$-ray sources
than XBLs, observations in the VHE range will help to confirm the
fundamental differences between the XBLs and RBLs. VHE $\gamma$-ray
observations of BL Lacs have, in general, concentrated on the closest
objects, but we have sought to extend the current redshift limit of $z =
0.117$ by observing more distant AGNs. With an energy threshold of $\sim
300$ GeV, the Mark 6 Telescope is well-suited to this task. In the case
of one XBL, 1ES 1101--232, the VHE $\gamma$-ray observations were made nearly
contemporaneously with {\it BeppoSAX} observations.

A total of 54 hours of on-source observations under clear skies of 7
objects was completed, and an observing log is shown in Table
\ref{observing_log}.

\section{Results}

\begin{table}[t!]
\begin{center}
\begin{tabular}{@{}lcclcc} \hline \hline
Object & Date & No. of &Object & Date & No. of\\
& & ON source scans & & & ON source scans \\ \hline
Cen A & 1997 Mar 08 & 5 & 1ES 2316--423 & 1997 Aug 30 & 13 \\
Cen A & 1997 Mar 10 & 5 & 1ES 2316--423 & 1997 Aug 03 & 11 \\
Cen A & 1997 Mar 11 & 6 & 1ES 2316--423 & 1997 Sep 06 & 4 \\
Cen A & 1997 Mar 12 & 6 & 1ES 1101--232 & 1998 May 19 & 6 \\
Cen A & 1997 Mar 13 & 5 & 1ES 1101--232 & 1998 May 20 & 3 \\
PKS 0829+046 & 1996 Mar 15 & 3 & 1ES 1101--232 & 1998 May 21 & 7 \\
PKS 0829+046 & 1996 Mar 17 & 6 & 1ES 1101--232 & 1998 May 22 & 6 \\
PKS 0829+046 & 1996 Mar 18 & 7 & 1ES 1101--232 & 1998 May 23 & 4 \\
PKS 1514--24 & 1996 Apr 14 & 2 & 1ES 1101--232 & 1998 May 24 & 4 \\
PKS 1514--24 & 1996 Apr 15 & 6 & 1ES 1101--232 & 1998 May 25 & 8 \\
PKS 1514--24 & 1996 Apr 17 & 6 & 1ES 1101--232 & 1998 May 26 & 8 \\
PKS 1514--24 & 1996 Apr 18 & 7 & 1ES 1101--232 & 1998 May 27 & 6 \\
PKS 1514--24 & 1996 Apr 19 & 10 & RXJ 10578--2753 & 1996 Mar 20 & 7 \\
PKS 1514--24 & 1996 Apr 20 & 6 & RXJ 10578--2753 & 1996 Mar 21 & 6 \\
PKS 1514--24 & 1996 Apr 21 & 6 & RXJ 10578--2753 & 1996 Mar 22 & 2 \\
PKS 1514--24 & 1996 Apr 22 & 8 & 1ES 0323+022 & 1996 Sep 14 & 4 \\
1ES 2316--423 & 1997 Aug 26 & 2 & 1ES 0323+022 & 1996 Sep 15 & 4 \\
1ES 2316--423 & 1997 Aug 27 & 2 & 1ES 0323+022 & 1996 Sep 17 & 7 \\
1ES 2316--423 & 1997 Aug 29 & 8 & & & \\ \hline
\end{tabular}
\end{center}
\caption{Observing log for observations of active galactic nuclei made
with the University of Durham Mark 6 Telescope.}
\label{observing_log} 
\end{table}

The flux limits from the seven AGNs are summarised in Table
\ref{results_table}. They are all $3\sigma$ flux limits, based on the
maximum likelihood ratio test (Gibson et al. 1982). The
threshold energy for the observations has been estimated on the basis of
preliminary simulations, and is in the range 300 to 400 GeV for these
objects, depending on the object's elevation.

\begin{wraptable}[14]{l}{10cm}
\begin{center}
\begin{tabular}{@{}lcc} \hline \hline
Object & Estimated & Flux Limit \\
& Threshold (GeV) & $ (\times 10^{-11} {\rm~cm}^{-2} {\rm~s}^{-1})$\\ \hline
Cen A & 300 & 5.2 \\
PKS 0829+046 & 400 & 4.7 \\
PKS 1514--24 & 300 & 3.7 \\
1ES 2316--423 & 300 & 4.5 \\
1ES 1101--232 & 300 & 3.7 \\
RXJ 10578--275 & 300 & 8.2 \\
1ES 0323+022 & 400 & 3.7 \\ \hline
\end{tabular}
\end{center}
\caption{Flux limits for observations of active galactic nuclei made
with the University of Durham Mark 6 Telescope.}
\label{results_table} 
\end{wraptable}
The collecting areas which
have been assumed, again from simulations, are $ 5.5 \times 10^{8} ~{\rm
cm}^{2}$ at an energy threshold of 300 GeV and $ 1.0 \times 10^{9} ~{\rm
cm}^{2}$ at an energy threshold of 400 GeV. These are subject to
systematic errors estimated to be $\sim 50 \%$. We have assumed that our
current selection procedures retain $\sim 20 \%$ of the $\gamma$-ray
signal, which is subject to a systematic error of $\sim 60$ \%. We have
also searched our dataset for $\gamma$-ray emission on timescales of
$\sim 1$ day. The search for enhanced emission has been conducted by
calculating the on-source excess after the application of our selection
criteria for the pairs of on/off observations recorded during an
individual night. A typical observation comprising 6 on/off pairs of
observations (1.5 hours of on-source observations) yields a flux limit
of $\sim 1 \times 10^{-10} {\rm ~cm}^{-2} {\rm ~s}^{-1}$ at 300 GeV.
Conversely, had any of the objects on which we report here produced a
15-minute flare similar to that seen from Mrk 421 with the Whipple
telescope on 1996 May 7 (Gaidos et al. 1996), it would be detected with
the Mark 6 telescope at a significance of around $7~\sigma$. There is no
evidence for any flaring activity.

\subsection{Cen A}

Centaurus A (NGC 5128) is the closest radio-loud active galaxy, at a
distance of 5 Mpc ($z = 0.008$), and is often described as the prototype
Fanaroff-Riley Class I low luminosity radio galaxy. It was identified as
a TeV source in the early days of VHE gamma ray astronomy (Grindlay et
al. 1975), with a flux of $4.4 \pm 1.0 \times 10^{-11} {\rm ~cm}^{-2}
{\rm ~s}^{-1}$ at an energy threshold of 300 GeV when the object was in
an X-ray high state. Observations of Cen A were also made with the
University of Durham Mark 3 telescope which placed a $3 \sigma$ flux
limit of $7.8 \times 10^{-11} {\rm ~cm}^{-2} {\rm ~s}^{-1}$ at a similar
energy threshold (Carraminana et al. 1990). The X-ray state of Cen A at
the time of these observations was unknown. Observations of Cen A made
in 1995 March/April with the CANGAROO telescope have resulted in a $3
\sigma$ flux limit of $1.28 \times 10^{-11} {\rm ~cm}^{-2}{\rm ~s}^{-1}$
at $E > 1.5$ TeV for an extended source centred on Cen A (Rowell et al.
1999). EGRET observations have recently been used to identify Cen A as a
source of GeV gamma rays (Sreekumar et al. 1999), thus providing the
first evidence for emission in the 30 -- 10000 MeV energy range from a
source with a confirmed large-inclination jet.

The observations of Cen A made with the Mark 6 imaging telescope
reported here provide a $3 \sigma$ flux limit of $5.2 \times 10^{-11} {\rm
~cm}^{-2} {\rm ~s}^{-1}$. {\em RXTE} observations taken
contemporaneously with our data confirm that Cen A was in a low state in
1997 March. If, as seems to be the case in other AGNs, the X-ray and VHE
$\gamma$-ray emission from Cen A are correlated, then it may not be
surprising that we detected no VHE emission in 1997 March.

\subsection{PKS 0829+046}

PKS 0829+046 ($z = 0.18$) was detected with both {\em HEAO-1} as an
X-ray source and the EGRET instrument as a GeV $\gamma$-ray source
(Fichtel et al. 1994). It has a large radio flux and is therefore
classified as an RBL; this suggests it is unlikely to be a detectable
VHE $\gamma$-ray source. The present $3 \sigma$ VHE limit is $4.7 \times
10^{-11} {\rm ~cm}^{-2} {\rm ~s}^{-1}$ for $E > 400$ GeV.

\subsection{PKS 1514--24}

PKS 1514--24 was one of the first radio-detected BL Lacs. It has a
redshift of 0.049, and although detected by {\em EXOSAT}, its relatively
small X-ray flux classifies it as an RBL. Phase 1 observations with the
EGRET detector on board {\em CGRO} resulted in an upper limit for the
object of $7 \times 10^{-8} {\rm ~cm}^{-2} {\rm ~s}^{-1}$ at $E > 100$
MeV (Fichtel et al. 1994) nor does it appear in the 3rd EGRET catalog
(Hartman et al. 1999). The VHE limit presented here is $3.7 \times
10^{-11} {\rm ~cm}^{-2}{\rm ~s}^{-1}$ for $E > 300$ GeV.

\subsection{1ES 2316--423}

1ES 2316--423 ($z = 0.055$) was originally classified as a radio
selected BL Lac, but recently Perlman et al. (1998) have identified this
object as an intermediate case whose high energy emission could be
expected to extend up to VHE energies. The CANGAROO telescope has
observed this object but detected no VHE emission, placing a $2 \sigma$
upper limit of $1.2 \times 10^{-12} {\rm~cm}^{-2} {\rm~s}^{-1}$ at a
threshold energy of $\sim 2$ TeV in July 1996 (Roberts et al. 1998). The
present measurement indicates a flux limit at the $3 \sigma$ level of
$4.5 \times 10^{-11} {\rm ~cm}^{-2} {\rm ~s}^{-1}$ at $E > 300$ GeV.
Assuming an integral spectral index of $\sim 1.5$ (c.f. Mrk 421 and Mrk
501), this corresponds to a $3\sigma$ flux limit at $E > 2$ TeV of $\sim
2.6 \times 10^{-12} {\rm ~cm}^{-2} {\rm ~s}^{-1}$, comparable with the
$2~\sigma$ flux limit from the CANGAROO experiment.

\subsection{1ES 1101--232}

1ES 1101--232 is an XBL with a redshift of 0.186. It has been detected
using both the {\em HEAO-1} and {\em Einstein} satellites. Phase one
EGRET observations resulted in an upper limit of $6 \times 10^{-8} {\rm
~cm}^{-2} {\rm ~s}^{-1}$ at $E > 100$ MeV (Fichtel et al. 1994). It was
detected with the {\it BeppoSAX} satellite in 1997 (Wolter et al. 1998),
and our observations of this XBL were made near contemporaneously with a {\it
BeppoSAX} campaign on the object. Indications are that the X-ray flux
from 1ES 1101--232 was $\sim 30\%$ lower during these observations than in
1997 (Wolter, private communication). Our $3 \sigma$ flux limit is $3.7
\times 10^{-11} {\rm~cm}^{-2}{\rm~s}^{-1}$ at $E > 300$ GeV.

\subsection{RXJ 10578--275}

The {\em Rosat} source RXJ 10578--275 was initially identified as a
potential BL Lac from its optical characteristics. It has a redshift of
0.092 and is classified as an XBL. Our $3 \sigma$ flux limit is $8.2
\times 10^{-11} {\rm ~cm}^{-2}{\rm ~s}^{-1}$ at $E > 300$ GeV.

\subsection{1ES 0323+022}

1ES 0323+022 has been detected using both the {\em HEAO-1} and {\em
Einstein} satellites. It is an XBL with a redshift of 0.147, and has a
spectrum which is very similar to the archetypal XBL PKS 2155--304.
EGRET phase one observations resulted in an upper limit of $ 6 \times
10^{-8} {\rm ~cm}^{-2} {\rm ~s}^{-1}$ (Fichtel et al. 1994). Stecker, de
Jager, \& Salamon (1996) predict a flux at $E > 300$ GeV of $4.0 \times
10 ^{-12}{\rm ~cm}^{-2}{\rm ~s}^{-1}$. Our $3 \sigma$ flux limit of~$3.7
\times 10^{-11} {\rm~cm}^{-2}{\rm~s}^{-1}$ at $E > 300$ GeV is
considerably higher and so it is not in conflict.

\section{Discussion}

Whilst the interpretation of VHE upper limits from BL Lacs is
complicated by the lack of a complete theory of VHE $\gamma$-ray
emission from AGNs, Stecker, de Jager, \& Salamon (1996) have predicted
the TeV fluxes from a range of objects, one of which (1ES 0323+022), is
included in the present work. The expected fluxes from the other XBLs
included in this paper may be estimated on the basis of the work of
Stecker, de Jager, \& Salamon (1996) and Stecker (1998) using the simple
relation $\nu_{x}F_{x} \sim \nu_{\gamma}F_{\gamma}$ and the published
X-ray fluxes. We estimate that the 300 GeV fluxes of 1ES 1101--232, 1ES
2316--423 and RXJ 10578--275 would be $2.0 \times 10 ^{-11}{\rm
~cm}^{-2}{\rm ~s}^{-1}$, $1.5 \times 10 ^{-12}{\rm ~cm}^{-2}{\rm
~s}^{-1}$, and $3.3 \times 10 ^{-12}{\rm ~cm}^{-2}{\rm ~s}^{-1}$
respectively, taking into account photon-photon absorption using the
recent determination of $\gamma$-ray opacity by Stecker (1998). All
these suggested fluxes are lower than the flux limits reported here.
However, the lack of contemporaneous X-ray measurements in the case of
most of our observations limits the usefulness of these predictions and
emphasises the importance of simultaneous X-ray and $\gamma$-ray
observations and multiwavelength campaigns. In the case of the RBLs, an
extended observation of PKS 1514--24, a close RBL, lends support to the
suggestion that RBLs are not strong VHE $\gamma$-ray emitters. 

Our observations of Cen A were made when it was in an X-ray low state,
in contrast to the earlier VHE detection of Cen A reported by
Grindlay et al. (1975) which was made when Cen A was in X-ray outburst.
Further VHE $\gamma$-ray observations during an X-ray high state would
be desirable.

We are grateful to the UK Particle Physics and Astronomy Research
Council for support of the project. We would like to thank Anna Wolter
for providing us with information about {\it BeppoSAX} observations of
1ES 1101--232 in advance of publication. This paper uses quick look
results provided by the ASM/{\it RXTE} team.

\vspace{1ex}
\begin{center}
{\Large\bf References}
\end{center}
%
Carraminana, A., et al. 1990, A\&A, 228, 327\\
Catanese, M. A., et al. 1998, \apj, 501, 616\\
Chadwick, P. M., et al. 1999, \apj, 513, 161\\
Fichtel, C. E., et al. 1994, \apjs, 94, 551\\ 
Gaidos, J. A., et al. 1996, \nat, 383, 319\\
Ghisellini, G., et al. 1998, \mnras, 301, 451\\
Gibson, A. I., et al. 1982, Proc. Intl. Workshop on Very High Energy
Gamma Ray Astro., Bombay: Tata Institute, ed. P. V. Ramana Murthy \& T.
C. Weekes, p. 97\\
Grindlay, J. E., Helmken, H. F., Hanbury Brown, R., Davis, J., \& Allen,
L. R. 1975, \apjl, 197, L9\\
Hartman, R. C., et al. 1999, \apjs, in press\\
Kerrick, A. D., et al. 1995, \apj, 452, 588\\
Perlman, E. S., et al. 1998, \aj, 115, 1253\\
Punch, M., et al. 1992, \nat, 358, 477\\
Quinn, J., et al. 1996, \apjl, 456, L83\\
Roberts, M. D., et al. 1998, A\&A, 337, 25\\
Roberts, M. D., et al. 1999, A\&A, 343, 691\\
Rowell, G. P., et al., 1999, astro-ph/9901316\\
Sreekumar, P., Bertsch, D. L., Hartman, R. C., Nolan, P. L., \&
Thompson, D. J. 1999, astro-ph/9901277\\
Stecker, F. W. 1998, astro-ph/9812286\\
Stecker, F. W., \& de Jager, O. C. 1998, A\&A, 334, L85\\
Stecker, F. W., de Jager, O. C., \& Salamon, M. H. 1992, \apjl, 390, L49\\
Stecker, F. W., de Jager, O. C., \& Salamon, M. H. 1996, \apjl, 473, L75\\
Wolter, A., et al. 1998, A\&A, 335, 899\\

\end{document}